\documentclass[10pt,twocolumn,twoside]{IEEEtran}
\usepackage{calc}

\usepackage{multirow}
\usepackage{cite}
\usepackage{graphicx,subfigure}
\usepackage{psfrag}
\usepackage{amsmath,amssymb}
\usepackage{color}

\usepackage{amsmath,amssymb,eucal}
\usepackage{xifthen}
\usepackage{mathtools}
\usepackage{enumerate}
\usepackage{microtype}
\usepackage{xspace}
\usepackage{bm}
\usepackage[T1]{fontenc}
\usepackage{fancyhdr}
\usepackage{lastpage}

\newcommand{\mbs}[1]{\bm{#1}}
\newcommand{\vect}[1]{{\lowercase{\mbs{#1}}}}
\newcommand{\mat}[1]{{\uppercase{\mbs{#1}}}}

\newcommand{\T}{{\scriptscriptstyle\mathsf{T}}}
\renewcommand{\H}{{\scriptscriptstyle\mathsf{H}}}

\renewcommand{\Re}[1][]{\ifthenelse{\isempty{#1}}{\operatorname{Re}}{\operatorname{Re}\left(#1\right)}}
\renewcommand{\Im}[1][]{\ifthenelse{\isempty{#1}}{\operatorname{Im}}{\operatorname{Im}\left(#1\right)}}

\newcommand{\gv}{\vect{g}}
\newcommand{\hv}{\vect{h}}

\newcommand{\uv}{\vect{u}}
\newcommand{\vv}{\vect{v}}
\newcommand{\wv}{\vect{w}}
\newcommand{\xv}{\vect{x}}
\newcommand{\yv}{\vect{y}}

\newcommand{\Hm}{\mat{h}}

\newcommand{\Qm}{\mat{q}}

\newcommand{\Xc}{{\mathcal X}}

\newcommand{\CC}{\mathbb{C}}

\newcommand{\CN}[1][]{\ifthenelse{\isempty{#1}}{\mathcal{N}_{\mathbb{C}}}{\mathcal{N}_{\mathbb{C}}\left(#1\right)}}

\renewcommand{\P}[1][]{\ifthenelse{\isempty{#1}}{\mathbb{P}}{\mathbb{P}\left(#1\right)}}
\newcommand{\E}[1][]{\ifthenelse{\isempty{#1}}{\mathbb{E}}{\mathbb{E}\left(#1\right)}}
\renewcommand{\det}[1][]{\ifthenelse{\isempty{#1}}{\text{det}}{\text{det}\left(#1\right)}}
\newcommand{\trace}[1][]{\ifthenelse{\isempty{#1}}{\text{tr}}{\text{tr}\left(#1\right)}}
\newcommand{\rank}[1][]{\ifthenelse{\isempty{#1}}{\text{rank}}{\text{rank}\left(#1\right)}}
\newcommand{\diag}[1][]{\ifthenelse{\isempty{#1}}{\text{diag}}{\text{diag}\left(#1\right)}}

\DeclarePairedDelimiter\norm{\lVert}{\rVert}

\newcommand{\defeq}{\triangleq}

\newtheorem{theorem}{Theorem}

\newcounter{enumi_saved}
\usepackage{answers}
\Newassociation{solution}{Solution}{solutionfile}

\AtBeginDocument{\Opensolutionfile{solutionfile}[\jobname]}
\AtEndDocument{\Closesolutionfile{solutionfile}\clearpage
}

\IfFileExists{MinionPro.sty}{
}{
}

\begin{document}
\sloppy

\title{Degrees-of-Freedom Region of the MISO Broadcast Channel with General Mixed-CSIT}
\author{Jinyuan Chen and Petros Elia
\thanks{The research leading to these results has received funding from the European Research Council under the European Community's Seventh Framework Programme (FP7/2007-2013) / ERC grant agreement no. 257616 (CONECT), from the FP7 CELTIC SPECTRA project, and from Agence Nationale de la Recherche project ANR-IMAGENET.
}
\thanks{J. Chen and P. Elia are with the Mobile Communications Department, EURECOM, Sophia Antipolis, France (email: \{chenji, elia\}@eurecom.fr)}
\thanks{This paper was submitted in part to the Information Theory Workshop (ITW) 2012.}
}


\maketitle
\thispagestyle{empty}


\begin{abstract}
In the setting of the two-user broadcast channel, recent work by Maddah-Ali and Tse has shown that knowledge of prior channel state information at the transmitter (CSIT) can be useful, even in the absence of any knowledge of current CSIT.  Very recent work by Kobayashi et al., Yang et al., and Gou and Jafar, extended this to the case where, instead of no current CSIT knowledge, the transmitter has partial knowledge, and where under a symmetry assumption, the quality of this knowledge is identical for the different users' channels.

Motivated by the fact that in multiuser settings, the quality of CSIT feedback may vary across different links, we here generalize the above results to the natural setting where the current CSIT quality varies for different users' channels.  For this setting we derive the optimal degrees-of-freedom (DoF) region, and provide novel multi-phase broadcast schemes that achieve this optimal region.  Finally this generalization incorporates and generalizes the corresponding result in Maleki et al. which considered the broadcast channel with one user having perfect CSIT and the other only having prior CSIT.

\end{abstract}

\section{Introduction}

In many multiuser wireless communications scenarios, having sufficient CSIT is a crucial ingredient that facilitates improved performance.  While being useful, perfect CSIT is also hard and time-consuming to obtain, hence the need for communication schemes that can utilize partial or delayed CSIT knowledge {(see \cite{DCSIT:GMKISIT11,DCSIT:VMKC10,DCSIT:AGKISIT11,DCSIT:GMK11,DCSIT:XAJ11})}.  In this context of multiuser communications, we here consider the broadcast channel (BC), and specifically focus on the two-user multiple-input single-output (MISO) BC, where a two-antenna transmitter communicates to two single-antenna receivers.  In this setting, the channel model takes the form
\begin{subequations}
\begin{align}
y^{(1)}_t &= \hv^{\T}_t \xv_t + z^{(1)}_t      \label{eq:modely1}\\
y^{(2)}_t &= \gv^{\T}_t \xv_t + z^{(2)}_t,      \label{eq:modely2}
\end{align}
\end{subequations}
where for any time instant $t$, $\hv_t, \gv_t \in \CC^{2\times 1}$ represent
the channel vectors for user~1 and 2 respectively, where $z^{(1)}_t,z^{(2)}_t$ represent unit power AWGN noise, where $\xv_t$ is the input signal with power
constraint $\E[ \norm{\xv_t}^2 ] \le P$, and where in this case, $P$ also takes the role of the signal-to-noise ratio (SNR).
It is well known that in this setting, the presence of full CSIT allows for the optimal $1$ degree-of-freedom (DoF) per user, whereas the complete absence of CSIT causes a substantial degradation to just $1/2$ DoF per user\footnote{We remind the reader that for an achievable rate pair $(R_1,R_2)$, the corresponding DoF pair $(d_1,d_2)$ is given by $d_i = \lim_{P \to \infty} \frac{R_i}{\log P},\ i=1,2.$  The corresponding DoF region is then the set of all achievable DoF pairs.}.

An interesting scheme that bridges this performance gap by utilizing partial CSIT knowledge, was recently presented in \cite{maddah2010degrees} which showed that delayed CSIT knowledge can still be useful in improving the DoF region of the broadcast channel.  In the above described two-user MISO BC setting, and under the assumption that at time $t$, the transmitter knows the delayed channel states ($\hv,\gv$) up to time $t-1$, the work in~\cite{maddah2010degrees} showed that each user can achieve $2/3$ DoF, providing a clear improvement over the case of no CSIT.

This result was later generalized in \cite{Kobayashi2012miso,Yang2012miso,Guo2012miso} which considered the natural extension where, in addition to the aforementioned perfect knowledge of prior CSIT, the transmitter also had imperfect knowledge of current CSIT; at time $t$ the transmitter had estimates $\hat{\hv}_t,\hat{\gv}_t$ of $\hv_t$ and $\gv_t$, with estimation errors
\begin{equation}
\label{MMSE}
  \tilde{\hv}_t  = \hv_t - \hat{\hv}_t, \ \ \     \tilde{\gv}_t  = \gv_t - \hat{\gv}_t
\end{equation}
having i.i.d. Gaussian entries with power \[\frac{1}{2}\E[\|\tilde{\hv}_t\|^2] = \frac{1}{2}\E[\|\tilde{\gv}_t\|^2] = P^{-\alpha},\] for some non-negative parameter $\alpha$ that described the quality of the estimate of the current CSIT.
In this setting of `mixed' CSIT (perfect prior CSIT and imperfect current CSIT), and for $d_1,d_2$ denoting the DoF for the first and second user over the aforementioned two-user BC, the work in \cite{Kobayashi2012miso,Yang2012miso,Guo2012miso}
showed the optimal DoF region to take the form,
\begin{equation}\label{eq:mixed1}\{d_1 \le 1; \ d_2 \le 1;  \ 2 d_1 + d_2 \le 2+\alpha; \ 2 d_2 + d_1 \le 2+\alpha\}\end{equation} corresponding to a polygon with corner points $\{(0,0), (1,0), (1, \alpha), (\frac{2+\alpha}{3}, \frac{2+\alpha}{3}), (\alpha, 1), (0, 1)\}$, nicely bridging the gap between the case of $\alpha = 0$ explored in \cite{maddah2010degrees}, and the case of $\alpha = 1$ (and naturally $\alpha > 1$) corresponding to perfect CSIT.

\subsection{Notation and conventions}
Throughout this paper, $(\bullet)^{-1}$, $(\bullet)^\T$, $(\bullet)^{\H}$, respectively denote the inverse, transpose, and conjugate transpose of a matrix, while $(\bullet)^{\ast}$ denotes the complex conjugate, and $||\bullet||$ denotes the Euclidean norm.
$|\bullet|$ denotes the magnitude of a scalar, and $\diag(\bullet)$ denotes a diagonal matrix. Logarithms are of base 2.
$o(\bullet)$ comes from the standard Landau notation, where $f(x) = o(g(x))$ implies $\lim_{x\to \infty} f(x)/g(x)=0$.  We also use $\doteq$ to denote \emph{exponential equality}, i.e., we write $f(P)\doteq P^{B}$ to denote $\displaystyle\lim_{P\to\infty}\frac{\log f(P)}{\log P}=B$.
Finally, in the spirit of \cite{Kobayashi2012miso,Yang2012miso,Guo2012miso} we consider a unit coherence period, as well as perfect knowledge of channel state information at the receivers (perfect CSIR).

\section{The generalized mixed-CSIT broadcast channel \label{sec:model-gmcsit}}

Motivated by the fact that in multiuser settings, the quality of CSIT feedback may vary across different links, we extend the approach in \cite{Kobayashi2012miso,Yang2012miso,Guo2012miso} to consider unequal quality of current CSIT knowledge for $\hv_t$ and $\gv_t$.
Specifically under the same set of assumptions mentioned above, and in the presence of perfect prior CSIT, we now consider the case where at time $t$, the transmitter has estimates $\hat{\hv}_t,\hat{\gv}_t$ of the current $\hv_t$ and $\gv_t$, with estimation errors
\begin{equation}
\label{MMSE2}
  \tilde{\hv}_t  = \hv_t - \hat{\hv}_t, \ \ \     \tilde{\gv}_t  = \gv_t - \hat{\gv}_t
\end{equation}
having i.i.d. Gaussian entries with power \[\frac{1}{2}\E[\|\tilde{\hv}_t\|^2] = P^{-\alpha_1}, \ \  \frac{1}{2}\E[\|\tilde{\gv}_t\|^2] = P^{-\alpha_2},\] for some non-negative parameters $\alpha_1,\alpha_2$ that describe the generally unequal quality of the estimates of the current CSIT for the two users' links.

We proceed to describe the optimal DoF region of the general mixed-CSIT two-user MISO BC (two-antenna transmitter).  The optimal schemes are presented in Section~\ref{sec:achievability}, parts of the proof of the schemes' performance are presented in Appendix~\ref{sec:Achievable}, while the outer bound proof is placed in Appendix~\ref{sec:outerb}.

\subsection{DoF region of the MISO BC with generalized mixed-CSIT  \label{sec:bc}}

Without loss of generality, the rest of this work assumes that
\begin{align}
1\geq \alpha_1 \geq \alpha_2 \geq 0.   \label{eq:assumption-alpha12}
\end{align}

\begin{theorem} \label{theorem:bc-outerb-gen}
The DoF region of the two-user MISO BC with general mixed-CSIT, is given by
  \begin{subequations}
  \begin{align}
    d_1 \le 1, \quad \  \  d_2 \le 1 \label{eq:them-outerb-d2}\\
    2 d_1  + d_2 \le 2+\alpha_1 \label{eq:them-outerb-2d1d2}\\
    d_1  + 2 d_2 \le 2+\alpha_2 \label{eq:them-outerb-2d2d1}
  \end{align}
  \end{subequations}
where the region is a polygon which, for $2\alpha_1-\alpha_2< 1$ has corner points \[\{\!(0,\!0),\! (1,\!0), \!(1, \!\alpha_1),\!(\frac{2\!+\!2\alpha_1\!-\!\alpha_2}{3},\! \frac{2\!+\!2\alpha_2\!-\!\alpha_1}{3}),\!(\alpha_2,\! 1), \!(0,\! 1)\!\},\] and otherwise
has corner points
\[\{(0,0), (1,0), (1, \frac{1+\alpha_2}{2}),(\alpha_2, 1), (0, 1)\}.\]
\end{theorem}
\vspace{3pt}

The above corner points, and consequently the entire DoF inner bound, will be attained by the schemes to be described later on.  The result generalizes the results in \cite{Kobayashi2012miso,Yang2012miso,Guo2012miso} as well as the result in \cite{Maleki2012ria} which considered the case of ($\alpha_1 = 1, \alpha_2 = 0$), where one user had perfect CSIT and the other only prior CSIT.

\begin{figure}
\centering
\includegraphics[width=9cm]{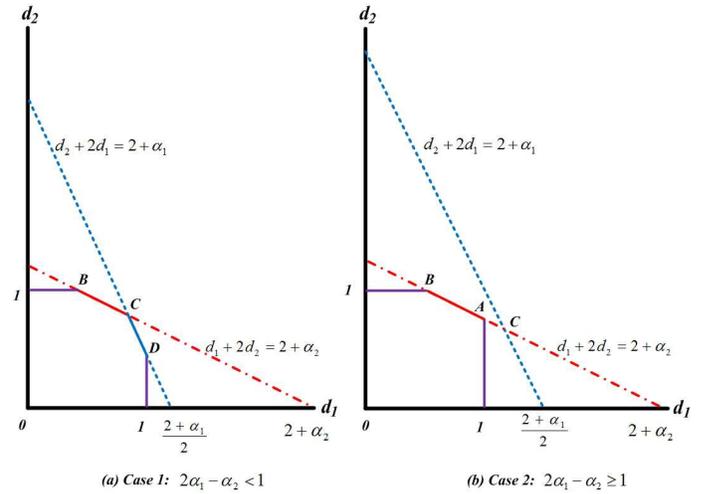}
\caption{DoF region when $2\alpha_1-\alpha_2< 1$ (case~1) and when $2\alpha_1-\alpha_2\geq 1$ (case~2).  The corner points take the following values: $A=(1, \frac{1+\alpha_2}{2})$, $B=(\alpha_2, 1)$, $C=(\frac{2+2\alpha_1-\alpha_2}{3}, \frac{2+2\alpha_2-\alpha_1}{3})$ and $D=(1, \alpha_1)$. }
\label{fig:DoFCSITOuterB}
\end{figure}

Figure~\ref{fig:DoFCSITOuterB} depicts the general DoF region for the case where $2\alpha_1-\alpha_2< 1$ (case~1) and the case where $2\alpha_1-\alpha_2\geq 1$ (case~2).

We proceed to describe the communication schemes.

\section{Design of communication schemes for the two-user general mixed-CSIT MISO BC} \label{sec:achievability}

As stated, without loss of generality, we assume that $1\geq \alpha_1 \geq \alpha_2 \geq 0$.
We describe the three schemes $\Xc_1$, $\Xc_2$ and $\Xc_3$ that achieve the optimal DoF region (in conjunction with time-division between these same schemes).  Specifically scheme $\Xc_1$ achieves $C=(\frac{2+2\alpha_1-\alpha_2}{3}, \frac{2+2\alpha_2-\alpha_1}{3})$ (case~1), scheme $\Xc_2$ achieves DoF points $D=(1, \alpha_1)$ (case~1) and $A=(1, \frac{1+\alpha_2}{2})$ (case~2), and scheme $\Xc_3$ achieves $B=(\alpha_2, 1)$ (case~1 and case~2).
The scheme description is done for $1>\alpha_1>\alpha_2\geq 0$, and for rational $\alpha_1,\alpha_2$.  The cases where $\alpha_1=1$, or $\alpha_1=\alpha_2$, or where $\alpha_1,\alpha_2$ are not rational, can be readily handled with minor modifications.
We proceed to describe the basic notation and conventions used in our schemes.

The schemes are designed with $S$ phases ($S$ varies from scheme to scheme), where the $s$th phase consists of $T_s$ channel uses, $s=1,2,\cdots,S$. The vectors $\hv_{s,t}$ and $\gv_{s,t}$ will denote the channel vectors seen by the first and second user respectively during timeslot $t$ of phase $s$, while $\hat{\hv}_{s,t}$ and $\hat{\gv}_{s,t}$ will denote the estimates of these channels at the transmitter during the same time, and $\tilde{\hv}_{s,t} = \hv_{s,t} - \hat{\hv}_{s,t} $, $\tilde{\gv}_{s,t} = \gv_{s,t} - \hat{\gv}_{s,t} $ will denote the estimation errors.

Furthermore $a_{s,t}$ and $a^{'}_{s,t}$ will denote the independent information symbols that may be sent during phase-$s$, timeslot-$t$, and which are meant for user 1, while symbols $b_{s,t}$ and $b^{'}_{s,t}$ are meant for user 2.
Vectors $\uv_{s,t}$ and $\vv_{s,t}$ are the unit-norm beamformers for $a_{s,t}$ and $b_{s,t}$ respectively, chosen so that $\uv_{s,t}$ is orthogonal to $\hat{\gv}_{s,t}$, and so that $\vv_{s,t}$ is orthogonal to $\hat{\hv}_{s,t}$. Furthermore $\uv^{'}_{s,t}, \vv^{'}_{s,t}$ are the randomly chosen unit-norm beamformers for $a^{'}_{s,t}$ and $b^{'}_{s,t}$ respectively.

Another notation that will be shared between schemes includes
\begin{align} \label{eq:cbar}
\bar{c}^{(b)}_{s,t} & \defeq \tilde{\hv}^\T_{s,t}\!\vv_{s,t} b_{s,t}\!+ \!\!\hv^\T_{s,t}\! \vv^{'}_{s,t} b^{'}_{s,t}, \nonumber\\
\bar{c}^{(a)}_{s,t} & \defeq \tilde{\gv}^\T_{s,t}\!\uv_{s,t} a_{s,t}\!+\! \!\gv^\T_{s,t} \!\uv^{'}_{s,t} a^{'}_{s,t}, \ t = 1,\cdots,T_s
\end{align}
that denotes the interference seen by user~1 and user~2 respectively, during timeslot $t$ of phase $s$.
For $\{\bar{c}^{(a)}_{s,t},\bar{c}^{(b)}_{s,t}\}_{t=1}^{T_{s}}$ being the accumulated interference to both users during phase $s$, we will let $\{\hat{c}^{(a)}_{s,t},\hat{c}^{(b)}_{s,t}\}_{t=1}^{T_{s}}$ be a quantized version of $\{\bar{c}^{(a)}_{s,t},\bar{c}^{(b)}_{s,t}\}_{t=1}^{T_{s}}$, and we will consider the mapping where the total information in $\{\hat{c}^{(a)}_{s,t},\hat{c}^{(b)}_{s,t}\}_{t=1}^{T_{s}}$ is split evenly across symbols $\{c_{s+1,t}\}_{t=1}^{T_{s+1}}$ transmitted during the next phase.
In addition we use $\wv_{s+1,t}$ to denote the randomly chosen unit-norm beamformer of $c_{s+1,t}$.

Furthermore, unless stated otherwise,
\begin{equation}\label{eq:TxGeneral}
\xv_{s,t} =\wv_{s,t} \underbrace{c_{s,t}}_{P_s^{(c)}}+\uv_{s,t} \underbrace{a_{s,t}}_{P_s^{(a)}} + \uv^{'}_{s,t} \underbrace{a'_{s,t}}_{P_s^{(a')}}+\vv_{s,t} \underbrace{b_{s,t}}_{P_s^{(b)}}+\vv^{'}_{s,t} \underbrace{b'_{s,t}}_{P_s^{(b')}}\end{equation} will be the general form of the transmitted vector at timeslot $t$ of phase $s$. 
As noted above under each summand, the average power that is assigned to each symbol, throughout a specific phase, will be denoted as follows: 
\[\begin{array}{ccc} P^{(c)}_s \defeq \E |c_{s,t}|^2, & P^{(a)}_s \defeq \E |a_{s,t}|^2, & P^{(a')}_s \defeq \E |a^{'}_{s,t}|^2 \\ P^{(b)}_s \defeq \E |b_{s,t}|^2, & P^{(b')}_s \defeq \E |b^{'}_{s,t}|^2.\end{array}\]
Furthermore each of the above symbols carries a certain amount of information, per timeslot, where this amount may vary across different phases. Specifically we use $r^{(a)}_s$ to mean that, during phase~$s$, each symbol $a_{s,t},  \ t = 1,\cdots,T_s,$ carries $r^{(a)}_s\log P +o(\log P)$ bits.  Similarly we use $r^{(a')}_s,r^{(b)}_s,r^{(b')}_s,r^{(c)}_s$ to describe the prelog factor of the number of bits in $a^{'}_{s,t}, b_{s,t},b^{'}_{s,t},c_{s,t}$ respectively, again for phase~$s$.

Finally the received signals during phase $s$ for the first and second user, are respectively denoted as $y^{(1)}_{s,t}$ and $y^{(2)}_{s,t}$, where generally the signals take the following form
\begin{align}
  y^{(1)}_{s,t}&= \hv^\T_{s,t} x_{s,t}+z^{(1)}_{s,t},\nonumber \\
  y^{(2)}_{s,t}&= \gv^\T_{s,t} x_{s,t}+z^{(2)}_{s,t}, \ t = 1,\cdots, T_s.
\end{align}


\subsection{Scheme $\Xc_1$ achieving $C=(\frac{2+2\alpha_1-\alpha_2}{3}, \frac{2+2\alpha_2-\alpha_1}{3})$ (case~1)}

As stated, scheme $\Xc_1$ has $S$ phases, where the phase durations $T_1,T_2,\cdots,T_S$ are chosen to be integers such that
\begin{align}
T_2&=T_1 \xi , \quad  T_s\!=\!T_{s-1} \mu\!=\!T_1\xi \mu^{s-2}, \forall s\in \{3,4,\cdots,S\!-\!1\}, \nonumber\\
T_{S}&=T_{S-1}\gamma=T_1\xi \mu^{S-3}\gamma,  \label{eq:sch2T}
\end{align}
where  $\xi=\frac{2-\alpha_1-\alpha_2}{1-\alpha_1-\Delta}$, $\
\mu=\frac{\alpha_1-\alpha_2+2\Delta}{1-\alpha_1-\Delta}$, $\gamma=\frac{\alpha_1-\alpha_2+2\Delta}{1-\alpha_2}$, and where $\Delta$ is any constant such that $0<\Delta<\frac{1-2\alpha_1+\alpha_2}{3}$.



\subsubsection{Phase~1}
During phase~1 ($T_1$ channel uses), the transmit signal is
\begin{equation} 
\label{eq:TxSch1Ph1}
\xv_{1,t} \!=\!\uv_{1,t} a_{1,t} \!+\! \uv^{'}_{1,t} a^{'}_{1,t}\!+\!\vv_{1,t} b_{1,t}\!+\!\vv^{'}_{1,t} b^{'}_{1,t}, \end{equation}
while the power and rate are set as
\begin{equation}\label{eq:ratePowerSch1Ph1}
\begin{array}{cccc}
P^{(a)}_1 \doteq P, & P^{(a')}_1 \doteq P^{1-\alpha_2}, & P^{(b)}_1 \doteq P, & P^{(b')}_1 \doteq P^{1-\alpha_1}\\
r^{(a)}_1  = 1, & r^{(a')}_1 = 1-\alpha_2, &  r^{(b)}_1 =1, & \ r^{(b')}_1 = 1-\alpha_1. \end{array} \end{equation}
The received signals at the two users then take the form
\begin{align}
  y^{(1)}_{1,t} &= \!\!\underbrace{\hv^\T_{1,t} \!\uv_{1,t} a_{1,t}}_{P} \!+\!\!\underbrace{\hv^\T_{1,t}\! \uv^{'}_{1,t} a^{'}_{1,t}}_{P^{1-\alpha_2}} \!+\!\!\overbrace{\underbrace{\tilde{\hv}^\T_{1,t}\! \vv_{1,t} b_{1,t}}_{P^{1-\alpha_1}}\! +\!\!\underbrace{\hv^\T_{1,t} \!\vv^{'}_{1,t} b^{'}_{1,t}}_{P^{1-\alpha_1}}}^{\bar{c}^{(b)}_{1,t}}\!+\!\!\underbrace{z^{(1)}_{1,t}}_{P^0}, \nonumber \\
  y^{(2)}_{1,t} &=\!\! \overbrace{\underbrace{\tilde{\gv}^\T_{1,t}\!\uv_{1,t} a_{1,t}}_{P^{1-\alpha_2}} \!+\!\!\underbrace{\gv^\T_{1,t} \!\uv^{'}_{1,t} a^{'}_{1,t}}_{P^{1-\alpha_2}}}^{\bar{c}^{(a)}_{1,t}} \!+\!\!\underbrace{\gv^\T_{1,t}\! \vv_{1,t} b_{1,t}}_{P}  \! +\!\!\underbrace{\gv^\T_{1,t} \!\vv^{'}_{1,t} b^{'}_{1,t}}_{P^{1-\alpha_1}}\!+\!\underbrace{z^{(2)}_{1,t}}_{P^0}, \label{eq:sch2y2}
\end{align}
where under each term we noted the order of the summand's average power.

At this point, and after the end of the first phase, the transmitter can use its knowledge of delayed CSIT to reconstruct $\{\bar{c}^{(a)}_{1,t}, \bar{c}^{(b)}_{1,t}\}_{t=1}^{T_1}$ (cf.\eqref{eq:cbar}), and quantize each term as
\begin{align}
  \bar{c}^{(a)}_{1,t} \!=\! \hat{c}^{(a)}_{1,t} \!+\! \tilde{c}^{(a)}_{1,t} , \quad \bar{c}^{(b)}_{1,t} \!=\! \hat{c}^{(b)}_{1,t} \!+\! \tilde{c}^{(b)}_{1,t} , \quad t=1,2,\cdots, T_1, \nonumber
\end{align}
where $\hat{c}^{(a)}_{1,t},\hat{c}^{(b)}_{1,t}$ are the quantized values, and where $\tilde{c}^{(a)}_{1,t},\tilde{c}^{(b)}_{1,t}$ are the quantization errors.
Noting that $\E|\bar{c}^{(a)}_{1,t}|^2 \doteq P^{1-\alpha_2}, \ \E|\bar{c}^{(b)}_{1,t}|^2 \doteq P^{1-\alpha_1}$, we choose a quantization rate that assigns each $\hat{c}^{(a)}_{1,t}$ a total of $(1-\alpha_2)\log P + o(\log P)$ bits, and each $\hat{c}^{(b)}_{1,t}$ a total of $(1-\alpha_1)\log P + o(\log P)$ bits, thus allowing for $\E|\tilde{c}^{(a)}_{1,t}|^2 \doteq \E|\tilde{c}^{(b)}_{1,t}|^2 \doteq 1$ (\cite{CT:06}).
At this point the $T_1(2-\alpha_1-\alpha_2)\log P+o(\log P)$ bits representing $\{\hat{c}^{(a)}_{1,t},\hat{c}^{(b)}_{1,t}\}_{t=1}^{T_1}$, are distributed evenly across the set $\{c_{2,t}\}_{t=1}^{T_2}$ which will be sequentially transmitted during the next phase. This transmission of $\{c_{2,t}\}_{t=1}^{T_2}$ will help each of the users cancel the interference from the other user, and it will also serve as an extra observation that allows for decoding of all private information of that same user.

\subsubsection{Phase~2}
During phase~2 ($T_2$ channel uses), the transmit signal takes the exact form in~\eqref{eq:TxGeneral}
\begin{equation}
\label{eq:TxSch1Ph2}\xv_{2,t} =\wv_{2,t} c_{2,t}+\uv_{2,t} a_{2,t} + \uv^{'}_{2,t} a^{'}_{2,t}+\vv_{2,t} b_{2,t}+\vv^{'}_{2,t} b^{'}_{2,t}\end{equation}
where we set power and rate as
\begin{equation}\label{eq:ratePowerSch1Ph2}
\begin{array}{ll}
P^{(c)}_2 \doteq P, & r^{(c)}_2  = 1-\alpha_1-\Delta \\
P^{(a)}_2 \doteq P^{\alpha_1+\Delta} , & r^{(a)}_2  = \alpha_1+\Delta\\
P^{(a')}_2 \doteq P^{\alpha_1-\alpha_2+\Delta} , & r^{(a')}_2 = \alpha_1-\alpha_2+\Delta\\
P^{(b)}_2 \doteq P^{\alpha_1+\Delta} , &  r^{(b)}_2 =\alpha_1+\Delta\\
P^{(b')}_2 \doteq P^{\Delta} , & \ r^{(b')}_2 = \Delta, \end{array} \end{equation}
and where we note that $r^{(c)}_2$ satisfies $T_2 r^{(c)}_2   = T_1(2-\alpha_1-\alpha_2)$.

The received signals during this phase are given as
\begin{align}
  y^{(1)}_{2,t}&=\! \underbrace{\hv^\T_{2,t} \wv_{2,t} c_{2,t}}_{P} \!+\!\underbrace{\hv^\T_{2,t} \uv_{2,t} a_{2,t}}_{P^{\alpha_1+\Delta}} \!+\!\underbrace{\hv^\T_{2,t} \uv^{'}_{2,t} a^{'}_{2,t}}_{P^{\alpha_1-\alpha_2+\Delta}} \nonumber\\
				&\! \quad +\!\underbrace{\tilde{\hv}^\T_{2,t} \vv_{2,t} b_{2,t}}_{P^{\Delta}}\!+\!\underbrace{\hv^\T_{2,t} \vv^{'}_{2,t} b^{'}_{2,t}}_{P^{\Delta}}\!+\!\underbrace{z^{(1)}_{2,t}}_{P^0}, \label{eq:sch2p2y1}\\
  y^{(2)}_{2,t}	&=\! \underbrace{\gv^\T_{2,t} \wv_{2,t} c_{2,t}}_{P} \!+\!\underbrace{\tilde{\gv}^\T_{2,t}\uv_{2,t} a_{2,t}}_{P^{\alpha_1-\alpha_2+\Delta}} \!+\!\underbrace{\gv^\T_{2,t} \uv^{'}_{2,t} a^{'}_{2,t}}_{P^{\alpha_1-\alpha_2+\Delta}} \nonumber\\
				&\!\quad +\underbrace{\gv^\T_{2,t} \vv_{2,t} b_{2,t}}_{P^{\alpha_1+\Delta}}+\underbrace{\gv^\T_{2,t} \vv^{'}_{2,t} b^{'}_{2,t}}_{P^{\Delta}}+\underbrace{z^{(2)}_{2,t}}_{P^0}, \label{eq:sch2p2y2}
\end{align}
for $ t\!=\!1,2,\!\cdots\!,\! T_2$, where under each term we noted the order of the summand's average power.

At this point, based on \eqref{eq:sch2p2y1},\eqref{eq:sch2p2y2}, each user decodes $c_{2,t}$ by treating the other signals as noise.  After decoding $\{c_{2,t}\}_{t=1}^{T_2}$ and fully reconstructing $\{\hat{c}^{(a)}_{1,t}, \hat{c}^{(b)}_{1,t},\}_{t=1}^{T_1}$, user 1 goes back one phase and subtracts $\hat{c}^{(b)}_{1,t}$ from $y^{(1)}_{1,t}$ to remove (up to bounded noise) the interference corresponding to $\bar{c}^{(b)}_{1,t}$.  The same user will also use the estimate $\hat{c}^{(a)}_{1,t}$ of $\bar{c}^{(a)}_{1,t}$ as an extra observation which, together with the observation $y^{(1)}_{1,t}$, present the user with a $2\times 2$ MIMO channel that allows for decoding of both $a_{1,t}$ and $a^{'}_{1,t}.$  Similarly user 2, after fully reconstructing $\{\hat{c}^{(a)}_{1,t}, \hat{c}^{(b)}_{1,t},\}_{t=1}^{T_1}$, subtracts $\hat{c}^{(a)}_{1,t}$ from $y^{(2)}_{1,t}$, to remove (up to bounded noise) the interference corresponding to $\bar{c}^{(a)}_{1,t}$, and also uses the estimate $\hat{c}^{(b)}_{1,t}$ of $\bar{c}^{(b)}_{1,t}$ as an extra observation which, together with the observation $y^{(2)}_{1,t}$, allow for decoding of both $b_{1,t}$ and $b^{'}_{1,t}.$  Further exposition to the details regarding the achievability of the mentioned rates, can be found in Appendix~\ref{sec:Achievable}.

Consequently after the end of the second phase, the transmitter can use its knowledge of delayed CSIT to reconstruct $\{\bar{c}^{(a)}_{2,t}, \bar{c}^{(b)}_{2,t}\}_{t=1}^{T_2}$, and quantize each term to $\hat{c}^{(a)}_{2,t},\hat{c}^{(b)}_{2,t}$.  With $\E|\bar{c}^{(a)}_{2,t}|^2 \doteq P^{\alpha_1-\alpha_2+\Delta}, \ \E|\bar{c}^{(b)}_{2,t}|^2 \doteq P^{\Delta}$, we choose a quantization rate that assigns each $\hat{c}^{(a)}_{2,t}$ a total of $(\alpha_1-\alpha_2+\Delta)\log P + o(\log P)$ bits, and each $\hat{c}^{(b)}_{2,t}$ a total of $\Delta\log P + o(\log P)$ bits, thus allowing for $\E|\tilde{c}^{(a)}_{2,t}|^2 \doteq \E|\tilde{c}^{(b)}_{2,t}|^2 \doteq 1$.
Then the $T_2(\alpha_1-\alpha_2+2\Delta)\log P+o(\log P)$ bits representing $\{\hat{c}^{(a)}_{2,t},\hat{c}^{(b)}_{2,t}\}_{t=1}^{T_2}$, are split evenly across the set $\{c_{3,t}\}_{t=1}^{T_3}$ which will be sequentially transmitted in the next phase so that user~1 can eventually decode $\{a_{2,t},a^{'}_{2,t}\}_{t=1}^{T_2}$, and user~2 can decode $\{b_{2,t},b^{'}_{2,t}\}_{t=1}^{T_2}$.

\begin{figure}
\centering
\includegraphics[width=8.5cm]{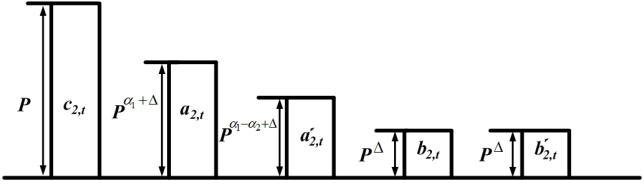}
\caption{Received power levels at user~1 (phase~2). }
\label{fig:X1P2R1}
\end{figure}

We now proceed with the general description of phase $s$.

\subsubsection{Phase~$s$, \ $3\leq s\leq S-1$}

Phase~$s$ ($T_s = T_{s-1} \frac{\alpha_1-\alpha_2+2\Delta}{1-\alpha_1-\Delta} $ channel uses) is almost identical to phase 2, with one difference being the different relationship between $T_s$ and $T_{s-1}$.  The transmit signal takes the same form as in phase 2 (cf.~\eqref{eq:TxGeneral},\eqref{eq:TxSch1Ph2}), the rates and powers of the symbols are the same (cf.~\eqref{eq:ratePowerSch1Ph2}) and the received signals $y^{(1)}_{s,t},y^{(2)}_{s,t}$ ($ t=1,\cdots,T_s$) take the same form as in \eqref{eq:sch2p2y1},\eqref{eq:sch2p2y2}.

Most of the actions are also the same, where based on \eqref{eq:sch2p2y1},\eqref{eq:sch2p2y2} (corresponding now to phase $s$), each user decodes $c_{s,t}$ by treating the other signals as noise, and then goes back one phase and reconstructs $\{\hat{c}^{(a)}_{s-1,t},\hat{c}^{(b)}_{s-1,t},\}_{t=1}^{T_{s-1}}$.  As before, user~1 then subtracts $\hat{c}^{(b)}_{s-1,t}$ from $y^{(1)}_{s-1,t}$ to remove, up to bounded noise, the interference corresponding to $\bar{c}^{(b)}_{s-1,t}$.  The same user also employs the estimate $\hat{c}^{(a)}_{s-1,t}$ of $\bar{c}^{(a)}_{s-1,t}$ as an extra observation which, together with the observation $y^{(1)}_{s-1,t}-\hv^\T_{s-1,t} \wv_{s-1,t} c_{s-1,t} - \hat{c}^{(b)}_{s-1,t}$ obtained after decoding $c_{s-1,t}$, allow for decoding of both $a_{s-1,t}$ and $a^{'}_{s-1,t}$.  Similar actions are performed by user~2.

As before, after the end of phase $s$, the transmitter can use its knowledge of delayed CSIT to reconstruct $\{\bar{c}^{(a)}_{s,t}, \bar{c}^{(b)}_{s,t}\}_{t=1}^{T_s}$, and quantize each term to $\hat{c}^{(a)}_{s,t},\hat{c}^{(b)}_{s,t}$ with the same rate as in phase~2 ($(\alpha_1-\alpha_2+\Delta)\log P + o(\log P)$ bits for each $\hat{c}^{(a)}_{s,t}$, and $\Delta\log P + o(\log P)$ bits for each $\hat{c}^{(b)}_{s,t}$). Finally the accumulated $T_{s}(\alpha_1-\alpha_2+2\Delta)\log P+o(\log P)$ bits representing all the quantized values $\{\hat{c}^{(a)}_{s,t},\hat{c}^{(b)}_{s,t}\}_{t=1}^{T_s}$, are distributed evenly across the set $\{c_{s+1,t}\}_{t=1}^{T_{s+1}}$ which will be sequentially transmitted in the next phase.
More details can be found in Appendix~\ref{sec:Achievable}.

\subsubsection{Phase~$S$}
During the last phase ($T_S = T_{S-1}\frac{\alpha_1-\alpha_2+2\Delta}{1-\alpha_2} $ channel uses), the transmit signal is
\begin{equation}\label{eq:TxSch1PhS}\xv_{S,t} =\wv_{S,t} c_{S,t}+\uv_{S,t} a_{S,t} + \vv_{S,t} b_{S,t}\end{equation}
where we set power and rate as
\begin{equation}\label{eq:ratePowerSch1PhS}
\begin{array}{ll}
P^{(c)}_S \doteq P, & r^{(c)}_S  = 1-\alpha_2 \\
P^{(a)}_S \doteq P^{\alpha_2} , & r^{(a)}_S  = \alpha_2\\
P^{(b)}_S \doteq P^{\alpha_2} , &  r^{(b)}_S =\alpha_2.
\end{array} \end{equation}
The received signals are
\begin{align}\label{eq:RxSch1PhS}
  y^{(1)}_{S,t}\!\!	&=\!\! \underbrace{\hv^\T_{S,t} \wv_{S,t} c_{S,t}}_{P} \!+\!\underbrace{\hv^\T_{S,t} \uv_{S,t} a_{S,t}}_{P^{\alpha_2}} \! +\!\underbrace{\tilde{\hv}^\T_{S,t} \vv_{S,t} b_{S,t}}_{P^{\alpha_2-\alpha_1}}\!+\!\underbrace{z^{(1)}_{S,t}}_{P^0}, \nonumber\\
  y^{(2)}_{S,t}\!\!&= \!\!\underbrace{\gv^\T_{S,t} \wv_{S,t} c_{S,t}}_{P} +\underbrace{\tilde{\gv}^\T_{S,t}\uv_{S,t} a_{S,t}}_{P^{0}} \! +\!\underbrace{\gv^\T_{S,t} \vv_{S,t} b_{S,t}}_{P^{\alpha_2}}+\underbrace{z^{(2)}_{S,t}}_{P^0}, 
\end{align}
for $ t\!=\!1,2,\!\cdots\!,T_S$.

At this point, as before, the power and rate allocation of the different symbols allow both users to decode $c_{S,t}$ by treating the other signals as noise.
Consequently user~1 can remove $\hv^\T_{S,t} \wv_{S,t} c_{S,t}$ from $y^{(1)}_{S,t}$ and decode $a_{S,t}$, and similarly user~2 can remove $\gv^\T_{S,t} \wv_{S,t} c_{S,t}$ from $y^{(2)}_{S,t}$ and decode $b_{S,t}$. Finally each user goes back one phase and reconstructs $\{\hat{c}^{(a)}_{S-1,t},\hat{c}^{(b)}_{S-1,t},\}_{t=1}^{T_{S-1}}$, which allows for decoding of $a_{S-1,t}$ and $a^{'}_{S-1,t}$ at user 1 and of $b_{S-1,t}$ and $b^{'}_{S-1,t}$ at user 2, all as described for the previous phases (see Appendix~\ref{sec:Achievable} for more details).

Table~\ref{tab:x1summary} summarizes the parameters of scheme $\Xc_1$.  The use of symbol $\bot$ is meant to indicate precoding that is orthogonal to the channel estimate (rather than random).  The table's last row indicates the prelog factor of the quantization rate.
\begin{table}[h]
\caption{Summary of scheme $\Xc_1$.}
\begin{center}
\begin{tabular}{|c|c|c|c|c|}
  \hline
                &Phase 1   &  Phase 2 & Ph. $s$ $(3\!\!\leq\!\! s \!\!\leq\!\! S\!\!-\!\!1)$& Phase $S$\\
   \hline
   Duration     &$T_1$ & $T_1\xi $ & $ T_1\xi \mu^{s-2} $ & $ T_1\xi \mu^{S-3}\gamma$ \\
   \hline
   $r^{(a)}$     &$1 $  & $ \alpha_1\!+\!\Delta$ & $ \alpha_1\!+\!\Delta$ & $ \alpha_2$ \\
    \hline
   $r^{(a')}$  &$1\!-\!\alpha_2 $ & $\alpha_1\!\!-\!\!\alpha_2\!\!+\!\!\Delta$ & $\alpha_1\!\!-\!\!\alpha_2\!\!+\!\!\Delta$ & - \\
    \hline
   $r^{(b)}$     &$1 $ & $ \alpha_1\!+\!\Delta$ & $ \alpha_1\!+\!\Delta$ & $ \alpha_2$ \\
    \hline
   $r^{(b')}$  &$1\!-\!\alpha_1 $ & $\Delta$ & $\Delta$ & - \\
    \hline
	 $r^{(c)}$     &-& $1\!-\!\alpha_1\!-\!\Delta$ & $1\!-\!\alpha_1\!-\!\Delta$ &  $1\!-\!\alpha_2$ \\
    \hline
   $P^{(a)}\bot$ 	 &$P$ & $P^{\alpha_1\!+\!\Delta}$ & $P^{\alpha_1\!+\!\Delta}$ & $P^{\alpha_2}$ \\
    \hline
   $P^{(a')}$  &$P^{1-\alpha_2}$ & $P^{\alpha_1\!-\!\alpha_2\!+\!\Delta}$ & $P^{\alpha_1\!-\!\alpha_2\!+\!\Delta}$ & - \\
    \hline
	 $P^{(b)}\bot$  &$P$ & $P^{\alpha_1\!+\!\Delta}$ & $P^{\alpha_1\!+\!\Delta}$ & $P^{\alpha_2}$ \\
    \hline
   $P^{(b')}$  &$P^{1-\alpha_1}$ & $P^{\Delta}$ & $P^{\Delta}$ & - \\
     \hline
   $P^{(c)}$      & - & $P$ & $P$ & $P$ \\
    \hline
   Quant.    &$2\!\!-\!\!\alpha_1\!\!\!-\!\!\alpha_2 $ & $\alpha_1\!\!\!-\!\!\alpha_2\!\!+\!\!2\!\Delta $ & $\alpha_1\!\!\!-\!\!\alpha_2\!\!+\!\!2\!\Delta $ & $0$ \\
    \hline
\end{tabular}
\end{center}
\label{tab:x1summary}
\end{table}

\paragraph{DoF calculation for scheme $\Xc_1$}
We proceed to add up the total amount of information transmitted during this scheme.  

In accordance to the declared pre-log factors $r_s^{(a)},r_s^{(a^{'})}$ and phase durations (see Table~\ref{tab:x1summary}), we have that
\begin{align}
d_1\!&=\!(T_1(2\!-\!\alpha_2)\!+\!\sum^{S-1}_{i=2}T_i(2\alpha_1\!-\!\alpha_2\!+\!2\Delta)\!+\!T_S\alpha_2)/(\sum^{S}_{i=1}T_i) \nonumber\\
&=( \sum^{S-1}_{i=2} (T_i(1\!-\!\alpha_1\!-\!\Delta)\!+\! T_i(\alpha_1\!+\!\Delta))  \!+\!T_{S}(1\!-\!\alpha_2)     \nonumber\\
&\quad \!+\!T_S\alpha_2 +T_1\alpha_1-\Delta\sum^{S-1}_{i=2}T_i  )/(\sum^{S}_{i=1}T_i) \label{eq:sch2d1a}\\
&=(1-\Delta)+\frac{T_1(\alpha_1+\Delta-1)+ T_S\Delta }{\sum^{S}_{i=1}T_i},
\end{align}
where \eqref{eq:sch2d1a} considers the phase durations seen in~\eqref{eq:sch2T}. 
Considering that $0<\mu<1$ (see \eqref{eq:sch2T} for case~1), that $\sum^{S-3}_{i=0}\mu^{i}=\frac{1-\mu^{S-2}}{1-\mu}$, and given an asymptotically high $S$, we see that
\begin{align}
d_1&=(1-\Delta)+\frac{\frac{T_2}{\xi}(\alpha_1+\Delta-1)+  T_2\mu^{S-3}\gamma\Delta }{\frac{T_2}{\xi}+T_2(\frac{1}{1-\mu} + \mu^{S-3}(\gamma-\frac{\mu}{1-\mu}) )} \label{eq:sch2d1b}\\ 
& =  (1-\Delta)+\frac{\frac{1}{\xi}(\alpha_1+\Delta-1) }{\frac{1}{\xi}+\frac{1}{1-\mu} } \nonumber\\
&= (1-\Delta)-\frac{1+\alpha_2-2\alpha_1-3\Delta }{3}  =\frac{2+2\alpha_1-\alpha_2}{3}. \label{eq:sch2d1final}
\end{align}
Similarly, considering the values for $r_s^{(b)},r_s^{(b^{'})}$, we have that
\begin{align}
d_2\!&=\!\frac{T_1(2-\alpha_1)+\sum^{S-1}_{i=2}T_i(\alpha_1+2\Delta)+T_S\alpha_2}{\sum^{S}_{i=1}T_i} \nonumber\\
&=\!\alpha_1\!+\!2\Delta\!+\!\frac{T_1(2\!-\!2\alpha_1\!-\!2\Delta)\!+\! T_S(\alpha_2\!-\!\alpha_1\!-\!2\Delta) }{\sum^{S}_{i=1}T_i} \nonumber\\
&=\!\alpha_1\!+\!2\Delta\!+\!\frac{\frac{T_2}{\xi}(2\!-\!2\alpha_1\!-\!2\Delta)\!\!+\!\!  T_2\mu^{S-3}\gamma(\alpha_2-\alpha_1-2\Delta) }{\frac{T_2}{\xi}+T_2(\frac{1}{1-\mu} + \mu^{S-3}(\gamma-\frac{\mu}{1-\mu}) )} \nonumber \end{align}
which, in the high $S$ limit, gives
\begin{align}
d_2& = \alpha_1+2\Delta+\frac{\frac{1}{\xi}(2-2\alpha_1-2\Delta) }{\frac{1}{\xi}+\frac{1}{1-\mu} } \nonumber\\
&=\! \alpha_1\!+\!2\Delta\!+\!\frac{2(1\!+\!\alpha_2\!-\!2\alpha_1\!-\!3\Delta) }{3} \!=\!\frac{2\!+\!2\alpha_2\!-\!\alpha_1}{3}. \label{eq:sch2d2}
\end{align}

In conclusion, scheme $\Xc_1$ achieves DoF pair $C=(\frac{2+2\alpha_1-\alpha_2}{3},\frac{2+2\alpha_2-\alpha_1}{3})$ (case~1).

\subsection{Scheme $\Xc_2$ achieving $D=(1, \alpha_1)$ (case~1), and $A=(1, \frac{1+\alpha_2}{2})$ (case~2)}

Scheme $\Xc_2$ is designed with $S$ phases, with phase durations $T_1,T_2,\cdots,T_S$ chosen to be integers such that
\begin{align}
T_2&=T_1\tau , \quad T_s\!=\!T_{s-1} \beta\!=\!T_1\tau \beta^{s-2}, \forall s\in \!\{3,4,\cdots,S\!-\!1\},  \nonumber\\
T_{S}&=T_{S-1}\eta=T_1\tau \beta^{S-3}\eta,  \label{eq:sch1T}
\end{align}
where $\tau=\frac{1-\alpha_2}{1-\alpha_1}$, $\beta=\frac{\alpha_1-\alpha_2}{1-\alpha_1}$, $\eta=\frac{\alpha_1-\alpha_2}{1-\alpha_2}$.

The scheme is similar to $\Xc_1$, but with a different power and rate allocation, and a different input structure since now user~2 only receives a single private information symbol.  



\subsubsection{Phase~1}
During phase~1 ($T_1$ channel uses), the transmitter sends 
\[\xv_{1,t} =\uv_{1,t} a_{1,t} + \uv^{'}_{1,t} a^{'}_{1,t}+\vv_{1,t} b_{1,t}, \]
with power and rate set as
\[
\begin{array}{ccc}
P^{(a)}_1 \doteq P, & P^{(a')}_1 \doteq P^{1-\alpha_2}, & P^{(b)}_1 \doteq P^{\alpha_1}\\
r^{(a)}_1  = 1, & r^{(a')}_1 = 1-\alpha_2, &  r^{(b)}_1 =\alpha_1. \end{array} \]
The received signals take the form
\begin{subequations}
\begin{align}
  y^{(1)}_{1,t}&= \underbrace{\hv^\T_{1,t} \uv_{1,t} a_{1,t}}_{P} +\underbrace{\hv^\T_{1,t} \uv^{'}_{1,t} a^{'}_{1,t}}_{P^{1-\alpha_2}} +\underbrace{\tilde{\hv}^\T_{1,t} \vv_{1,t} b_{1,t}}_{P^{0}}+\underbrace{z^{(1)}_{1,t}}_{P^0}, \nonumber\\
  y^{(2)}_{1,t}	&= \overbrace{\underbrace{\tilde{\gv}^\T_{1,t}\uv_{1,t} a_{1,t}}_{P^{1-\alpha_2}} +\underbrace{\gv^\T_{1,t} \uv^{'}_{1,t} a^{'}_{1,t}}_{P^{1-\alpha_2}}}^{\bar{c}^{(a)}_{1,t}}  +\underbrace{\gv^\T_{1,t} \vv_{1,t} b_{1,t}}_{P^{\alpha_1}}+\underbrace{z^{(2)}_{1,t}}_{P^0}. \nonumber
\end{align}
\end{subequations}

After the end of the first phase, the transmitter reconstructs $\{\bar{c}^{(a)}_{1,t}\}_{t=1}^{T_1}$ (cf.\eqref{eq:cbar}), and quantizes each term as
\begin{align}
  \bar{c}^{(a)}_{1,t} \!=\! \hat{c}^{(a)}_{1,t} \!+\! \tilde{c}^{(a)}_{1,t} , \quad t=1,2,\cdots, T_1. \nonumber
\end{align}
Noting that $\E|\bar{c}^{(a)}_{1,t}|^2 \doteq P^{1-\alpha_2}$, we choose a quantization rate that assigns each $\hat{c}^{(a)}_{1,t}$ a total of $(1-\alpha_2)\log P + o(\log P)$ bits, thus allowing for $\E|\tilde{c}^{(a)}_{1,t}|^2 \doteq 1$.  Then the $T_1(1-\alpha_2)\log P+o(\log P)$ bits representing $\{\hat{c}^{(a)}_{1,t}\}_{t=1}^{T_1}$ are distributed evenly across the set $\{c_{2,t}\}_{t=1}^{T_2}$ which will be transmitted in the next phase.  As before, transmission of $\{c_{2,t}\}_{t=1}^{T_2}$ aims to help user~2 cancel out interference, as well as aims to provide user~1 with an extra observation which will allow for decoding of the user's private information.

\subsubsection{Phase~2}
During phase~2 ($T_2$ channel uses), the transmitter sends 
\[\xv_{2,t} =\wv_{2,t} c_{2,t}+\uv_{2,t} a_{2,t} + \uv^{'}_{2,t} a^{'}_{2,t}+\vv_{2,t} b_{2,t}\]
with power and rate set as
\begin{equation}\label{eq:ratePowerPhase2X2}
\begin{array}{ll}
P^{(c)}_2 \doteq P, & r^{(c)}_2  = 1-\alpha_1 \\
P^{(a)}_2 \doteq P^{\alpha_1} , & r^{(a)}_2  = \alpha_1\\
P^{(a')}_2 \doteq P^{\alpha_1-\alpha_2} , & r^{(a')}_2 = \alpha_1-\alpha_2\\
P^{(b)}_2 \doteq P^{\alpha_1} , &  r^{(b)}_2 =\alpha_1,
\end{array} \end{equation}
where we note that $r^{(c)}_2$ satisfies $T_2 r^{(c)}_2   = T_1(1-\alpha_2)$.

The received signals in this phase are 
\begin{align}
  y^{(1)}_{2,t}\!\!&=\!\! \underbrace{\hv^\T_{2,t} \!\wv_{2,t} c_{2,t}}_{P} \!+\!\underbrace{\hv^\T_{2,t} \!\uv_{2,t} a_{2,t}}_{P^{\alpha_1}} \! +\!\underbrace{\hv^\T_{2,t} \!\uv^{'}_{2,t} a^{'}_{2,t}}_{P^{\alpha_1-\alpha_2}} \!+\!\underbrace{\tilde{\hv}^\T_{2,t} \!\vv_{2,t}\! b_{2,t}}_{P^{0}}\!+\!\!\underbrace{z^{(1)}_{2,t}}_{P^0} \label{eq:sch1p2y1}\\
  y^{(2)}_{2,t}\!\!&=\!\! \underbrace{\gv^\T_{2,t} \!\wv_{2,t} c_{2,t}}_{P} +\underbrace{\tilde{\gv}^\T_{2,t}\!\uv_{2,t} a_{2,t}}_{P^{\alpha_1-\alpha_2}}\!+\!\underbrace{\gv^\T_{2,t} \!\uv^{'}_{2,t} a^{'}_{2,t}}_{P^{\alpha_1-\alpha_2}}\! +\!\underbrace{\gv^\T_{2,t} \!\vv_{2,t}\! b_{2,t}}_{P^{\alpha_1}}\!+\!\!\underbrace{z^{(2)}_{2,t}}_{P^0} \label{eq:sch1p2y2}
\end{align}
for $ t\!=\!1,2,\!\cdots\!,\! T_2$.

Then, based on \eqref{eq:sch1p2y1},\eqref{eq:sch1p2y2}, each user decodes $c_{2,t}$ by treating the other signals as noise, and then proceeds to reconstruct $\{\hat{c}^{(a)}_{1,t}\}_{t=1}^{T_1}$.  User~1 combines each $\hat{c}^{(a)}_{1,t}$ with its corresponding observation $y^{(1)}_{1,t}$, to introduce $T_2$ independent $2\times 2$ MIMO channels that allow for decoding of all $a_{1,t}$ and $a^{'}_{1,t}.$  At the same time, user~2 subtracts $\hat{c}^{(a)}_{1,t}$ from $y^{(2)}_{1,t}$ to remove (up to bounded noise) the interference corresponding to $\bar{c}^{(a)}_{1,t}$, which in turn allows for decoding of $b_{1,t}$.   

Consequently after the end of the second phase, the transmitter can use its knowledge of delayed CSIT to reconstruct $\{\bar{c}^{(a)}_{2,t}\}_{t=1}^{T_2}$, and quantize each term to $\hat{c}^{(a)}_{2,t}$.  With $\E|\bar{c}^{(a)}_{2,t}|^2 \doteq P^{\alpha_1-\alpha_2}$, we choose a quantization rate that assigns each $\hat{c}^{(a)}_{2,t}$ a total of $(\alpha_1-\alpha_2)\log P + o(\log P)$ bits, a choice that allows for $\E|\tilde{c}^{(a)}_{2,t}|^2 \doteq 1$.
Then the $T_2(\alpha_1-\alpha_2)\log P+o(\log P)$ bits representing $\{\hat{c}^{(a)}_{2,t}\}_{t=1}^{T_2}$, are distributed evenly across the set $\{c_{3,t}\}_{t=1}^{T_3}$ which will be transmitted in the next phase.

\subsubsection{Phase~$s$, \ $3\leq s\leq S-1$}

Phase~$s$ ($T_s = T_{s-1} \frac{\alpha_1-\alpha_2}{1-\alpha_1} $ channel uses) is almost identical to phase 2, except for the relationship between $T_s$ and $T_{s-1}$.
Specifically the transmit signal takes the same form as in phase 2 
\[
\xv_{s,t} =\wv_{s,t} \underbrace{c_{s,t}}_{P_s^{(c)}}+\uv_{s,t} \underbrace{a_{s,t}}_{P_s^{(a)}} + \uv^{'}_{s,t} \underbrace{a'_{s,t}}_{P_s^{(a')}}+\vv_{s,t} \underbrace{b_{s,t}}_{P_s^{(b)}},\]
the rates and powers of the symbols are the same (cf.~\eqref{eq:ratePowerPhase2X2}), and the received signals $y^{(1)}_{s,t},y^{(2)}_{s,t}$ ($ t=1,\cdots,T_s$) take the same form as in \eqref{eq:sch1p2y1},\eqref{eq:sch1p2y2}.

The actions are also the same, where based on \eqref{eq:sch1p2y1},\eqref{eq:sch1p2y2} (corresponding now to phase $s$), each user decodes $c_{s,t}$ by treating the other signals as noise, and then goes back one phase and reconstructs $\{\hat{c}^{(a)}_{s-1,t}\}_{t=1}^{T_{s-1}}$.  As before, user~1 then employs the estimate $\hat{c}^{(a)}_{s-1,t}$ of $\bar{c}^{(a)}_{s-1,t}$ as an extra observation which, together with the observation $y^{(1)}_{s-1,t} - \hv^\T_{s-1,t} \wv_{s-1,t} c_{s-1,t}$ attained after decoding $c_{s-1,t}$, allow for decoding of both $a_{s-1,t}$ and $a^{'}_{s-1,t}.$
At the same time, user~2 subtracts $\hat{c}^{(a)}_{s-1,t}$ from $y^{(2)}_{s-1,t}$ to remove (up to bounded noise) the interference corresponding to $\bar{c}^{(a)}_{s-1,t}$, which allows for decoding of $b_{s-1,t}$.

Again as before, after the end of phase~$s$, the transmitter can use delayed CSIT to reconstruct $\{\bar{c}^{(a)}_{s,t}\}_{t=1}^{T_s}$, and quantize each term to $\hat{c}^{(a)}_{s,t}$ with the same rate as in phase~2 ($(\alpha_1-\alpha_2)\log P + o(\log P)$ bits per channel use). Finally the total of the $T_{s}(\alpha_1-\alpha_2)\log P+o(\log P)$ bits representing the quantized values $\{\hat{c}^{(a)}_{s,t}\}_{t=1}^{T_s}$ is split evenly to the set $\{c_{s+1,t}\}_{t=1}^{T_{s+1}}$ which will be transmitted in the next phase.

\subsubsection{Phase~$S$}
During the last phase ($T_S = T_{S-1}\frac{\alpha_1-\alpha_2}{1-\alpha_2} $ channel uses), the transmitter sends
\begin{equation}\label{eq:transSch1PhS}\xv_{S,t} =\wv_{S,t} c_{S,t}+\uv_{S,t} a_{S,t} + \vv_{S,t} b_{S,t}\end{equation}
with power and rates set as
\begin{equation}\label{eq:ratePowerPhaseSX2}
\begin{array}{ll}
P^{(c)}_S \doteq P, & r^{(c)}_S  = 1-\alpha_2 \\
P^{(a)}_S \doteq P^{\alpha_2} , & r^{(a)}_S  = \alpha_2\\
P^{(b)}_S \doteq P^{\alpha_2} , &  r^{(b)}_S =\alpha_2.
\end{array} \end{equation}
resulting in received signals of the form
\begin{subequations}
\begin{align}
  y^{(1)}_{S,t}\!\!	&=\!\! \underbrace{\hv^\T_{S,t} \wv_{S,t} c_{S,t}}_{P} \!+\!\underbrace{\hv^\T_{S,t} \uv_{S,t} a_{S,t}}_{P^{\alpha_2}} \! +\!\underbrace{\tilde{\hv}^\T_{S,t} \vv_{S,t} b_{S,t}}_{P^{\alpha_2-\alpha_1}}\!+\!\underbrace{z^{(1)}_{S,t}}_{P^0}, \nonumber\\
  y^{(2)}_{S,t}\!\!&= \!\!\underbrace{\gv^\T_{S,t} \wv_{S,t} c_{S,t}}_{P} +\underbrace{\tilde{\gv}^\T_{S,t}\uv_{S,t} a_{S,t}}_{P^{0}} \! +\!\underbrace{\gv^\T_{S,t} \vv_{S,t} b_{S,t}}_{P^{\alpha_2}}+\underbrace{z^{(2)}_{S,t}}_{P^0}, \nonumber
\end{align}
\end{subequations}
($ t\!=\!1,\!\cdots\!,T_S$).

As before, both receivers decode $c_{S,t}$ by treating all other signals as noise.  Consequently user~1 removes $\hv^\T_{S,t} \wv_{S,t} c_{S,t}$ from $y^{(1)}_{S,t}$ and decodes $a_{S,t}$, and user~2 removes $\gv^\T_{S,t} \wv_{S,t} c_{S,t}$ from $y^{(2)}_{S,t}$ and decodes $b_{S,t}$. Finally each user goes back one phase and reconstructs $\{\hat{c}^{(a)}_{S-1,t}\}_{t=1}^{T_{S-1}}$, which in turn allows for decoding of $a_{S-1,t}$ and $a^{'}_{S-1,t}$ at user~1 and of $b_{S-1,t}$ at user~2, all as described in the previous phases.  The DoF achievability details follow those of scheme $\Xc_1$ (Appendix~\ref{sec:Achievable}).

Table~\ref{tab:x2summary} summarizes the parameters of scheme $\Xc_2$.  The last row indicates the prelog factor of the quantization rate.
\begin{table}[h]
\caption{Summary of scheme $\Xc_2$.}
\begin{center}
\begin{tabular}{|c|c|c|c|c|}
  \hline
                &Phase 1   &  Phase 2 & Ph.$s$ $(3\!\!\leq\!\! s \!\!\leq\!\! S\!\!-\!\!1)$& Phase $S$\\
   \hline
   Duration     &$T_1$ & $T_1\tau $ & $ T_1\tau \beta^{s-2} $ & $ T_1\tau \beta^{S-3}\eta$ \\
   \hline
   $r^{(a)}$     &$1 $  & $ \alpha_1$ & $ \alpha_1$ & $ \alpha_2$ \\
    \hline
   $r^{(a')}$  &$1\!-\!\alpha_2 $ & $\alpha_1\!\!-\!\!\alpha_2$ & $\alpha_1\!\!-\!\!\alpha_2$ & - \\
    \hline
   $r^{(b)}$     &$\alpha_1 $ & $ \alpha_1$ & $ \alpha_1$ & $ \alpha_2$ \\
    \hline
	 $r^{(c)}$     &-& $1\!-\!\alpha_1$ & $1\!-\!\alpha_1$ &  $1\!-\!\alpha_2$ \\
    \hline
   $P^{(a)}\bot$ &$P$ & $P^{\alpha_1}$ & $P^{\alpha_1}$ & $P^{\alpha_2}$ \\
    \hline
   $P^{(a')}$  &$P^{1-\alpha_2}$ & $P^{\alpha_1\!-\!\alpha_2}$ & $P^{\alpha_1\!-\!\alpha_2}$ & - \\
    \hline
	 $P^{(b)}\bot$ &$P^{\alpha_1}$ & $P^{\alpha_1}$ & $P^{\alpha_1}$ & $P^{\alpha_2}$ \\
    \hline
   $P^{(c)}$       & - & $P$ & $P$ & $P$ \\
    \hline
   Quant.    &$1\!-\!\alpha_2 $ & $\alpha_1\!-\!\alpha_2 $ & $\alpha_1\!-\!\alpha_2 $ & $0$ \\
    \hline
\end{tabular}
\end{center}
\label{tab:x2summary}
\end{table}

\paragraph{DoF calculation for scheme $\Xc_2$}
We proceed to add up the total amount of information transmitted during this scheme.

In accordance to the declared pre-log factors $r_s^{(a)},r_s^{(a^{'})}$ and phase durations (see Table~\ref{tab:x2summary}), and irrespective of whether $\alpha_1,\alpha_2$ fall under case~1 or case~2, we have that
\begin{align}
d_1\!\!&=\!\!(T_1(2\!-\!\alpha_2)\!+\!\sum^{S-1}_{i=2}T_i(2\alpha_1\!-\!\alpha_2)\!+\!T_S\alpha_2)/(\sum^{S}_{i=1}T_i) \nonumber\\
&=\!\!(T_1\!\!+\!T_1(1\!\!-\!\alpha_2)\!+\!\!\! \sum^{S-1}_{i=2} \!(T_i\alpha_1\!+\!T_i(\alpha_1\!\!-\!\alpha_2)) \!+\!T_S\alpha_2)/(\sum^{S}_{i=1}\!T_i)\nonumber\\
&=\!\!(T_1\!\!+\!\!\sum^{S-1}_{i=2} \!(T_i(1\!\!-\!\alpha_1)\!+\!T_i\alpha_1) \!+\!T_{S}(1\!\!-\!\alpha_2)\!+\!T_S\alpha_2)/(\sum^{S}_{i=1}T_i) \label{eq:sch3d1sp}\\
&=\!\!\frac{T_1+T_2+T_3+\cdots+T_{S-1}+T_S}{T_1+T_2+\cdots+T_S} =1 \label{eq:sch3d1final}
\end{align}
where \eqref{eq:sch3d1sp} is due to \eqref{eq:sch1T}.

Regarding the second user and the declared $r_s^{(b)}$, for case 1 ($2\alpha_1-\alpha_2<1$) we see that 
\begin{align}
d_2&=\frac{\sum^{S-1}_{i=1}T_i\alpha_1+T_S\alpha_2}{\sum^{S}_{i=1}T_i} =\alpha_1-\frac{T_S(\alpha_1-\alpha_2)}{\sum^{S}_{i=1}T_i} \nonumber\\
&=\alpha_1-\frac{T_1\tau \beta^{S-3}\eta(\alpha_1-\alpha_2)}{T_1+  T_1\tau\sum^{S-3}_{i=0}\beta^{i}+T_1\tau \beta^{S-3}\eta} \label{eq:sch3d2sp} \\
&=\alpha_1-\frac{ \beta^{S-3}\eta(\alpha_1-\alpha_2)}{\frac{1}{\tau}+  \sum^{S-3}_{i=0}\beta^{i}+ \beta^{S-3}\eta} \label{eq:sch3d2a}\\
&=\alpha_1-\frac{ \beta^{S-3}\eta(\alpha_1-\alpha_2)}{\frac{1}{\tau}+  \frac{1-\beta^{S-2}}{1-\beta}+ \beta^{S-3}\eta} \nonumber\\
&=\alpha_1-\frac{ \beta^{S-3}\eta(\alpha_1-\alpha_2)}{\frac{1}{\tau}+ \frac{1}{1-\beta}+ \beta^{S-3}(\eta-\frac{\beta}{1-\beta})}  =  \alpha_1, \label{eq:sch3d2case2}
\end{align}
where we have used~\eqref{eq:sch1T} to get~\eqref{eq:sch3d2sp}, where we have used that $2\alpha_1-\alpha_2<1$ implies $\beta<1$, and where we have considered an asymptotically large $S$.

When $2\alpha_1-\alpha_2>1$ ($\beta>1$), then \eqref{eq:sch3d2a} gives that 
\begin{align}
d_2&=\alpha_1-\frac{ \beta^{S-3}\eta(\alpha_1-\alpha_2)}{\frac{1}{\tau}+ \frac{1}{1-\beta}+ \beta^{S-3}(\eta-\frac{\beta}{1-\beta})} \nonumber\\
&=\alpha_1-\frac{ \eta(\alpha_1-\alpha_2)}{\frac{1-\beta+\tau}{\beta^{S-3}\tau(1-\beta)}+ (\eta-\frac{\beta}{1-\beta})} \nonumber
\end{align}
which, in the high $S$ regime, gives
\begin{align}
d_2& = \alpha_1 -\frac{ \eta(\alpha_1-\alpha_2)}{ \eta-\frac{\beta}{1-\beta}} \!=\! \alpha_1 \!+\!\frac{1\!-\!2\alpha_1\!+\!\alpha_2}{2}  \!=\! \frac{1\!+\!\alpha_2}{2}. \label{eq:sch3d2case1}
\end{align}

When $2\alpha_1-\alpha_2=1$ ($\beta=1$), then \eqref{eq:sch3d2a} gives that $d_2=\alpha_1-\frac{ \eta(\alpha_1-\alpha_2)}{\frac{1}{\tau}+  S-2+ \eta}$ which, for large $S$, gives 
\begin{align}
d_2= \alpha_1 = \frac{1+\alpha_2}{2}.
\end{align}

In conclusion, scheme $\Xc_2$ achieves DoF pair $D=(1, \alpha_1)$ (case~1), else it achieves $A=(1, \frac{1+\alpha_2}{2})$.

\subsection{Scheme $\Xc_3$ achieving $B=(\alpha_2, 1)$}

This is the simplest of all three schemes, and it consists of a single channel use\footnote{We will henceforth maintain the same notation as before, but for simplicity we will remove the phase and time index.} ($S = 1, T_1 = 1$) during which the transmitter sends
\[\xv =\wv c+ \uv a + \vv b,\]
where $\uv$ is orthogonal to $\hat{\gv}$, $\vv$ is orthogonal to $\hat{\hv}$, and where the power and rates are set as
\begin{equation}\label{eq:ratePowerPhaseSX3}
\begin{array}{ll}
P^{(c)} \doteq P, & r^{(c)}  = 1-\alpha_1 \\
P^{(a)} \doteq P^{\alpha_2} , & r^{(a)}  = \alpha_2\\
P^{(b)} \doteq P^{\alpha_1} , &  r^{(b)} =\alpha_1,
\end{array} \end{equation}
resulting in received signals of the form
\begin{subequations}
\begin{align}
  y^{(1)} &= \hv^\T \xv + z^{(1)}= \underbrace{\hv^\T \wv c}_{P} +\underbrace{\hv^\T \uv a}_{P^{\alpha_2}} +\underbrace{\tilde{\hv}^\T \vv b}_{P^0}+\underbrace{z^{(1)}}_{P^0},  \nonumber\\ 
  y^{(2)} &= \gv^\T \xv + z^{(2)}= \underbrace{\gv^\T \wv c}_{P} +\underbrace{\tilde{\gv}^\T \uv a}_{P^0} +\underbrace{\gv^\T \vv b}_{P^{\alpha_1}}+\underbrace{z^{(2)}}_{P^0}. \nonumber
\end{align}
\end{subequations}

After transmission, both receivers first decode $c$ by treating the other signals as noise, and then user~1 utilizes its knowledge of $\{\hv,\gv,\hat{\hv},\hat{\gv}\}$ to reconstruct $\hv^\T \wv c$ and remove it from $y^{(1)}$, thus being able to decode $a$, while after decoding $c$, user~2 removes $\gv^\T \wv c$ from $y^{(2)}$, and decodes $b$.
The details for the achievability of $r^{(a)},r^{(b)},r^{(c)}$ follow closely the exposition in Appendix~\ref{sec:Achievable}.  Consequently the DoF point $(d_1=\alpha_2, \quad d_2=1)$ can be achieved by associating $c$ to information intended entirely for the second user.

\section{Conclusions} \label{sec:conclu}

The work provided analysis and communication schemes for the setting of the two-user MISO BC with general mixed CSIT. The work can be seen as a natural extension of the result in \cite{Maleki2012ria} and of the recent results in \cite{maddah2010degrees,Kobayashi2012miso,Yang2012miso,Guo2012miso}, to the case where the CSIT feedback quality varies across different links.

\section{Appendix - Details of achievability proof} \label{sec:Achievable}

We will here focus on achievability details for scheme $\Xc_1$.  The clarifications of the details carry over easily to the other two schemes.

Regarding $r^{(c)}_s$ ($2\leq s \leq S-1$ - see~\eqref{eq:ratePowerSch1Ph2}), we recall that during phase~$s$, both users decode $c_{s,t}$ (from $y^{(1)}_{s,t}, y^{(2)}_{s,t}, t=1,\cdots,T_s$ - see \eqref{eq:sch1p2y1},\eqref{eq:sch1p2y2} ) by treating all other signals as noise.  Consequently for $\Hm\!\defeq \!\{ \hv_{i,j},  \gv_{i,j},\hat{\hv}_{i,j}, \hat{\gv}_{i,j}, \forall i,j\}$, we note that \begin{align}
I(c_{s,t};y^{(1)}_{s,t},\Hm)&=I(c_{s,t};y^{(2)}_{s,t},\Hm) \nonumber\\
&=(1-\alpha_1-\Delta)\log P+ o(\log P),  \nonumber
\end{align}
to get 
\begin{align}
r^{(c)}_s & = \frac{1}{\log P} \min \{I(c_{s,t};y^{(1)}_{s,t},\Hm),I(c_{s,t};y^{(2)}_{s,t},\Hm)\}  \nonumber\\
&=1-\alpha_1-\Delta.\nonumber
\end{align}
Similarly for the last phase~$S$ (see~\eqref{eq:TxSch1PhS},\eqref{eq:ratePowerSch1PhS},\eqref{eq:RxSch1PhS}), we note that
\begin{align}
I(c_{S,t};y^{(1)}_{S,t},\Hm)\!=\!I(c_{S,t};y^{(2)}_{S,t},\Hm)\!=\!(1\!-\!\alpha_2)\log P\!+\! o(\log P),  \nonumber
\end{align}
to get
\begin{align}
r^{(c)}_S &= \frac{1}{\log P} \min \{I(c_{S,t};y^{(1)}_{S,t},\Hm),I(c_{S,t};y^{(2)}_{S,t},\Hm)\} =1-\alpha_2.\nonumber
\end{align}

Regarding achievability for $r^{(a)}_1=1$, $r^{(a^{'})}_1=1-\alpha_2$, $r^{(b)}_1=1$ and $r^{(b^{'})}_1=1-\alpha_1$ (see~\eqref{eq:TxSch1Ph1},\eqref{eq:ratePowerSch1Ph1},\eqref{eq:sch2y2}), we note that each element in $\{c_{2,t}\}_{t=1}^{T_2}$ has enough bits (recall that $r^{(c)}_2=1-\alpha_1-\Delta$), to match the quantization rate of $\{\hat{c}^{(a)}_{1,t}, \hat{c}^{(b)}_{1,t}\}_{t=1}^{T_1}$ that is necessary in order to have a bounded quantization noise.  Consequently going back to phase~1, user~1 is presented with $T_1$ linearly independent $2\times 2$ equivalent MIMO channels of the form
\begin{align}
\begin{bmatrix} \!y^{(1)}_{1,t}\!-\!\hat{c}^{(b)}_{1,t}
          \\ \hat{c}^{(a)}_{1,t} \!\end{bmatrix}   \!\!=\!\! \begin{bmatrix} \! \hv^\T_{1,t} \\ \gv^\T_{1,t} \!\end{bmatrix} \!\!\begin{bmatrix} \! \uv_{1,t}  \   \uv^{'}_{1,t} \!\end{bmatrix} \!\!\begin{bmatrix} \!a_{1,t} \\  a^{'}_{1,t} \!\end{bmatrix}
 \!\!+\!\!\! {\begin{bmatrix} \!z^{(1)}_{1,t} \!+ \!\tilde{c}^{(b)}_{1,t}\\ -
              \tilde{c}^{(a)}_{1,t} \!\end{bmatrix}} \nonumber
\end{align}
($t=1,2,\cdots,T_1$), where again we note that the described quantization rate results in a bounded equivalent noise, which then immediately gives that $r^{(a)}_1=1$ and $r^{(a^{'})}_1=1-\alpha_2$ are achievable.  Similarly for user~2, the presented $T_1$ linearly independent $2\times 2$ equivalent MIMO channels 
\begin{align}
	\begin{bmatrix} \! \hat{c}^{(b)}_{1,t}
          \\ y^{(2)}_{1,t}\!-\! \hat{c}^{(a)}_{1,t} \!\end{bmatrix} \!\!=\!\! \begin{bmatrix} \! \hv^\T_{1,t} \\ \gv^\T_{1,t} \!\end{bmatrix} \!\!\begin{bmatrix} \! \vv_{1,t}  \   \vv^{'}_{1,t} \!\end{bmatrix} \!\!\begin{bmatrix} \!b_{1,t} \\  b^{'}_{1,t} \!\end{bmatrix}
 \!+\! {\begin{bmatrix} \! -\tilde{c}^{(b)}_{1,t}\\
              z^{(2)}_{1,t} \!+\! \tilde{c}^{(a)}_{1,t} \!\end{bmatrix}} \nonumber
\end{align}
($t=1,2,\cdots,T_1$), allow for decoding at a rate corresponding to $r^{(b)}_1=1$ and $r^{(b^{'})}_1=1-\alpha_1$.

Regarding achievability for $r^{(a)}_s=\alpha_1+\Delta$, $r^{(a^{'})}_s=\alpha_1-\alpha_2+\Delta$, $r^{(b)}_s=\alpha_1+\Delta$ and $r^{(b^{'})}_s=\Delta$, ($2\leq s\leq S-1$ - see~\eqref{eq:TxGeneral},\eqref{eq:TxSch1Ph2},~\eqref{eq:ratePowerSch1Ph2}), we note that during phase~$s$, both users can decode $c_{s,t}$, and as a result user~1 can remove $\hv^\T_{s,t}\wv_{s,t} c_{s,t}$ from $y^{(1)}_{s,t}$, and user~2 can remove $\gv^\T_{s,t}\wv_{s,t} c_{s,t}$ from $y^{(2)}_{s,t}$ ($t=1,\cdots,T_s$).  As a result user~1 is presented with $T_s$ linearly independent $2\times 2$ equivalent MIMO channels of the form
\begin{align}
\begin{bmatrix} \!y^{(1)}_{s,t}-\hv^\T_{s,t}\wv_{s,t} c_{s,t}\!-\!\hat{c}^{(b)}_{s,t}
          \\ \hat{c}^{(a)}_{s,t} \!\end{bmatrix}   \!\!=\!\! \begin{bmatrix} \! \hv^\T_{s,t} \\ \gv^\T_{s,t} \!\end{bmatrix} \!\!\begin{bmatrix} \! \uv_{s,t}  \   \uv^{'}_{s,t} \!\end{bmatrix} \!\!\begin{bmatrix} \!a_{s,t} \\  a^{'}_{s,t} \!\end{bmatrix}
 \!\!+\!\!\! {\begin{bmatrix} \!z^{(1)}_{s,t} \!+ \!\tilde{c}^{(b)}_{s,t}\\ -
              \tilde{c}^{(a)}_{s,t} \!\end{bmatrix}} \nonumber
\end{align}
($t=1,\cdots,T_s$).  Given that the rate associated to $\{c_{s+1,t}\}_{t=1}^{T_{s+1}}$, matches the quantization rate for $\{\hat{c}^{(a)}_{s,t}, \hat{c}^{(b)}_{s,t}\}_{t=1}^{T_s}$, allows for a bounded variance of the equivalent noise, and in turn for decoding of $\{a_{s,t},a^{'}_{s,t}\}_{t=1}^{T_s}$ at a rate corresponding to $r^{(a)}_s=\alpha_1+\Delta$ and $r^{(a^{'})}_s=\alpha_1-\alpha_2+\Delta$.
Similarly user~2 is presented with $T_s$ independent $2\times 2$ MIMO channels of the form 
\begin{align}
	\begin{bmatrix} \! \hat{c}^{(b)}_{s,t}
          \\ y^{(2)}_{s,t}-\gv^\T_{s,t}\wv_{s,t} c_{s,t}\!-\! \hat{c}^{(a)}_{s,t} \!\end{bmatrix} \!\!=\!\! \begin{bmatrix} \! \hv^\T_{s,t} \\ \gv^\T_{s,t} \!\end{bmatrix} \!\!\begin{bmatrix} \! \vv_{s,t}  \   \vv^{'}_{s,t} \!\end{bmatrix} \!\!\begin{bmatrix} \!b_{s,t} \\  b^{'}_{s,t} \!\end{bmatrix}
 \!+\! {\begin{bmatrix} \! -\tilde{c}^{(b)}_{s,t}\\
              z^{(2)}_{s,t} \!+\! \tilde{c}^{(a)}_{s,t} \!\end{bmatrix}} \nonumber
\end{align}
allowing for decoding of $\{b_{s,t},b^{'}_{s,t}\}_{t=1}^{T_s}$ ($t=1,\cdots,T_s$) at rates corresponding to $r^{(b)}_s=\alpha_1+\Delta$ and $r^{(b^{'})}_s=\Delta$.

Regarding achievability for $r^{(a)}_S=\alpha_2$ and $r^{(b)}_S=\alpha_2$ (see~\eqref{eq:TxSch1PhS},\eqref{eq:ratePowerSch1PhS},\eqref{eq:RxSch1PhS}), we note that, after decoding $c_{S,t}$, user~1 can remove $\hv^\T_{S,t} \wv_{S,t} c_{S,t}$ from $ y^{(1)}_{S,t}$, and user~2 can remove $\gv^\T_{S,t} \wv_{S,t} c_{S,t}$ from $y^{(2)}_{S,t}$, ($t=1,\cdots,T_S$).  Consequently during this phase, user~1 sees $T_S$ linearly independent SISO channels of the form
\begin{align}
\tilde{y}^{(1)}_{S,t}\!\defeq\! y^{(1)}_{S,t}\!-\!\hv^\T_{S,t} \wv_{S,t} c_{S,t}\!=\! \hv^\T_{S,t} \uv_{S,t} a_{S,t} \!+\!\tilde{\hv}^\T_{S,t} \vv_{S,t} b_{S,t}\!+\!z^{(1)}_{S,t} \nonumber
\end{align}
($t=1,\cdots,T_S$) which can be readily shown to support $r^{(a)}_S=\alpha_2$.  A similar argument gives achievability for $r^{(b)}_S=\alpha_2$.
\hfill $\Box$

\section{Appendix - Proof of Outer bound\label{sec:outerb}} 

We here adopt the outer bound approach in \cite{Guo2012miso} to the asymmetric case of $\alpha_1\neq \alpha_2$.
As in \cite{Guo2012miso}, we first linearly convert the original BC in~\eqref{eq:modely1},\eqref{eq:modely2} to an equivalent BC (see~\eqref{eq:model-ttr-y1},\eqref{eq:model-ttr-y2}) having the same DoF region as the original BC (cf.\cite{Guo2012miso}), and we then consider the degraded version of the equivalent BC in the absence of delayed feedback, which matches in capacity the degraded BC with feedback (for the memoryless case), and which exceeds the capacity of the equivalent BC.  The final step considers the compound and degraded version of the equivalent BC without delayed feedback, whose DoF region will serve as an outer bound on the DoF region of the original BC.

\paragraph{The equivalent degraded compound BC}

Towards the equivalent BC, directly from~\eqref{eq:modely1},\eqref{eq:modely2} we have that
\begin{subequations}
\begin{align}
y^{(1)}_{t} 	&= \hv^{\T}_{t}\xv_{t} + z^{(1)}_{t}    \nonumber \\
				&= \hv^{\T}_{t}\sqrt{P}\Qm_{t} \frac{1}{\sqrt{P}}\Qm^{-1}_{t}\xv_{t} + z^{(1)}_{t}   \nonumber \\
				&= \hv^{\T}_{t}\sqrt{P}\Qm_{t}\xv^{'}_{t} + z^{(1)}_{t}   \nonumber \\
				&= \sqrt{P}\hv^{\T}_{t}\uv_{t}x^{1}_{t}+ \sqrt{P}\tilde{\hv}^{\T}_{t}\vv_{t}x^{2}_{t} + z^{(1)}_{t}  	 \label{eq:model-ttr-y1}\\
y^{(2)}_{t} 	&= \gv^{\T}_{t}\xv_{t} + z^{(2)}_{t}    \nonumber \\
				&= \gv^{\T}_{t}\sqrt{P}\Qm_{t}\xv^{'}_{t} + z^{(2)}_{t}    \nonumber \\
				&= \sqrt{P}\tilde{\gv}^{\T}_{t}\uv_{t}x^{1}_{t}+ \sqrt{P}\gv^{\T}_{t}\vv_{t}x^{2}_{t} + z^{(2)}_{t},      \label{eq:model-ttr-y2}
\end{align}
\end{subequations}
where \[\xv^{'}_{t} \defeq [x^{1}_{t} \ x^{2}_{t}]^{T} \defeq \frac{1}{\sqrt{P}} \Qm^{-1}_{t} \xv_{t}, \] where $\Qm_{t}\defeq [\uv_{t} \ \vv_{t}] \in \mathbb{C}^{2\times 2}$ is, with probability 1, an invertible matrix, where $\uv_{t}$ is chosen to be of unit norm and orthogonal to $\hat{\gv}_{t}$, and where $\vv_{t}$ is chosen to be of unit norm and orthogonal to $\hat{\hv}_{t}$.  Furthermore each receiver normalizes to get
\begin{subequations}
\begin{align}
y^{'(1)}_{t}&=\frac{y^{(1)}_{t}}{\hv^{\T}_{t}\uv_{t}}   \nonumber \\
							&=\sqrt{P}x^{1}_{t}+ \frac{\sqrt{P}\tilde{\hv}^{\T}_{t}\vv_{t}x^{2}_{t}}{\hv^{\T}_{t}\uv_{t}} + \frac{z^{(1)}_{t}}{\hv^{\T}_{t}\uv_{t}}      \nonumber \\
							&=\sqrt{P}x^{1}_{t}+ \sqrt{P^{1-\alpha_1}}h^{'}_{t}x^{2}_{t} + z^{'(1)}_{t} ,     \label{eq:model-rtr-y1}\\
y^{'(2)}_{t}&=\frac{y^{(2)}_{t}}{\gv^{\T}_{t}\vv_{t}} \nonumber \\
							&=\sqrt{P}x^{2}_{t}+ \frac{\sqrt{P}\tilde{\gv}^{\T}_{t}\uv_{t}x^{1}_{t}}{\gv^{\T}_{t}\vv_{t}}  + \frac{z^{(2)}_{t}}{\gv^{\T}_{t}\vv_{t}}		 \nonumber \\
							&=\sqrt{P}x^{2}_{t}+ \sqrt{P^{1-\alpha_2}}g^{'}_{t}x^{1}_{t} + z^{'(2)}_{t} ,      \label{eq:model-rtr-y2}
\end{align}
\end{subequations}
where $z^{'(1)}_{t}=\frac{z^{(1)}_{t}}{\hv^{\T}_{t}\uv_{t}}$, $h^{'}_{t}=\frac{\sqrt{P^{\alpha_1}}\tilde{\hv}^{\T}_{t}\vv_{t}}{\hv^{\T}_{t}\uv_{t}}$, $z^{'(2)}_{t}=\frac{z^{(2)}_{t}}{\gv^{\T}_{t}\vv_{t}}$, $g^{'}_{t}=\frac{\sqrt{P^{\alpha_2}}\tilde{\gv}^{\T}_{t}\uv_{t}}{\gv^{\T}_{t}\vv_{t}}$.  Consequently $\sqrt{P^{\alpha_1}}\tilde{\hv}_{t}$ and $\sqrt{P^{\alpha_2}}\tilde{\gv}_{t}$ have identity covariance matrices, and the average power of $h^{'}_{t}$, $g^{'}_{t}$, $z^{'(1)}_{t}$ and $z^{'(2)}_{t}$ does not scale with $P$, i.e., in the high-SNR region this power is of order $P^{0}$.  With the same CSIT knowledge mapped from the original BC, it can be shown (see \cite{Guo2012miso}) that the DoF region of the equivalent BC in \eqref{eq:model-rtr-y1}\eqref{eq:model-rtr-y2} matches the DoF region of the original BC in \eqref{eq:modely1}\eqref{eq:modely2}.

Towards designing the degraded version of the above equivalent BC, we supply the second user with knowledge of $y^{'(1)}_{t}$, and towards designing the compound version of the above degraded equivalent BC, we add two extra users (user 3 and 4).  In this compound version, the received signals for the first two users are as in \eqref{eq:model-rtr-y1}\eqref{eq:model-rtr-y2}, while the received signals of the added (virtual) users are given by
\begin{subequations}
\begin{align}
y^{''(1)}_{t}&=\sqrt{P}x^{1}_{t}+ \sqrt{P^{1-\alpha_1}}h^{''}_{t}x^{2}_{t} + z^{''(1)}_{t} ,     \label{eq:model-comp-y1}\\
y^{''(2)}_{t}&=\sqrt{P}x^{2}_{t}+ \sqrt{P^{1-\alpha_2}}g^{''}_{t}x^{1}_{t} + z^{''(2)}_{t}.      \label{eq:model-comp-y2}
\end{align}
\end{subequations}
We here note that by definition, $h^{''}_{t}$ and $g^{''}_{t}$ are statistically equivalent to the original $h^{'}_{t}$ and $g^{'}_{t}$ respectively, and that $z^{''(1)}_{t}$ and $z^{''(2)}_{t}$ are statistically equivalent to the original $z^{'(1)}_{t}$ and $z^{'(2)}_{t}$.
Furthermore we note that user~3 is interested in the same message as user~1, while user~4 is interested in the same message as user~2.  Also we recall that in the specific degraded compound BC, user~1 knows $y^{'(1)}_{t}$, user~2 knows $y^{'(2)}_{t}$ and $y^{'(1)}_{t}$, user~3 knows $y^{''(1)}_{t}$, and user~4 knows $y^{''(2)}_{t}$ and $y^{''(1)}_{t}$.
Finally we remove delayed feedback - a removal known to not affect the capacity of the degraded BC without memory~\cite{Gamal1978feedback}.

We now proceed to calculate an outer bound on the DoF region of this degraded compound BC which at least matches the DoF of the previous degraded BC and which serves as an outer bound on the DoF region of the original BC.

\paragraph{Outer bound}
We consider communication over the described equivalent degraded compound BC, letting $n$ be the large number of fading realizations over which communication takes place, and letting $R_1,R_2$ be the rates of the first and second user. We also let
$\Hm_{[n]}\defeq \{\hv_{t}, \gv_{t},\hat{\hv}_{t}, \hat{\gv}_{t}\}^{n}_{t=1}$,
$y^{'(i)}_{[n]}\defeq \{y^{'(i)}_{t}\}_{t=1}^{n}$ and $y^{''(i)}_{[n]} \defeq \{y^{''(i)}_{t}\}_{t=1}^{n}$ for $i=1,2$.

Using Fano's inequality, we have
\begin{align}
nR_1 \!&\leq\! I(W_1;y^{'(1)}_{[n]}|\Hm_{[n]}) + n o(n)   \nonumber \\
&\leq\!  n \log\! P \!+\! n o(\log\! P) \!-\! h(y^{'(1)}_{[n]}|W_1,\Hm_{[n]}) \!+\! n o(n),   \label{eq:R1-bound-2}
\end{align}
as well as
\begin{align}
nR_1 \!&\leq\! I(W_1;y^{''(1)}_{[n]}|\Hm_{[n]}) + n o(n)  \nonumber \\
&\leq\! n \log \! P\!+\! n o(\log\! P)\!-\! h(y^{''(1)}_{[n]}|W_1,\Hm_{[n]}) \!+\! n o(n),   \label{eq:R1-bound-12}
\end{align}
which is added to \eqref{eq:R1-bound-2} to give
\begin{align}
2nR_1 &\leq 2n\log P+ 2n o(\log P)-h(y^{'(1)}_{[n]}|W_1,\Hm_{[n]}) \nonumber \\
&\quad  	-h(y^{''(1)}_{[n]}|W_1,\Hm_{[n]})+ 2n o(n) \nonumber \\
&\leq   	2n\log P+ 2n o(\log P) \nonumber \\
&\quad 		-h(y^{'(1)}_{[n]},y^{''(1)}_{[n]}|W_1,\Hm_{[n]}) + 2n o(n) .   \label{eq:2R1-bound}
\end{align}

Let
\begin{align}
\bar{\yv_1} &\defeq
\diag(1, \sqrt{P^{\alpha_1}})
{\begin{bmatrix}
			1  \quad h^{'}_{t} \\
			1  \quad  h^{''}_{t}	
\end{bmatrix} }^{-1}
\begin{bmatrix}
			y^{'(1)}_{t} \\
			y^{''(1)}_{t}	
\end{bmatrix}        \nonumber \\
&= \begin{bmatrix}
			\sqrt{P}x^{1}_{t} \\
			\sqrt{P}x^{2}_{t}
\end{bmatrix}
 + \begin{bmatrix}
			\frac{z^{'(1)}_{t}h^{''}_{t}- z^{''(1)}_{t}h^{'}_{t}}{h^{''}_{t}-h^{'}_{t}} \\
			 \sqrt{P^{\alpha_1}}\frac{z^{''(1)}_{t}- z^{'(1)}_{t}}{h^{''}_{t}-h^{'}_{t}}
\end{bmatrix} \nonumber \\
& = \begin{bmatrix}
			\sqrt{P}x^{1}_{t} \\
			\sqrt{P}x^{2}_{t}
\end{bmatrix}
 + \begin{bmatrix}
			\bar{z}_{t} \\
			 0
\end{bmatrix}
 + \begin{bmatrix}
			0 \\
			 z_{t}
\end{bmatrix}
\end{align}
where $\bar{z}_{t}=\frac{z^{'(1)}_{t}h^{''}_{t}- z^{''(1)}_{t}h^{'}_{t}}{h^{''}_{t}-h^{'}_{t}}$, $z_{t}=\sqrt{P^{\alpha_1}}\frac{z^{''(1)}_{t}- z^{'(1)}_{t}}{h^{''}_{t}-h^{'}_{t}}$, and let $z_{[n]}\defeq \{z_{t}\}_{t=1}^{n}$.  Consequently
\begin{align}
nR_1+nR_2 &= h(W_1,W_2)  \nonumber \\
					&= I(W_1,W_2;y^{'(1)}_{[n]},y^{''(1)}_{[n]},z_{[n]}|\Hm_{[n]}) \nonumber \\ &\quad  +h(W_1,W_2|y^{'(1)}_{[n]},y^{''(1)}_{[n]},z_{[n]},\Hm_{[n]})  \nonumber \\
					&= I(W_1,W_2;y^{'(1)}_{[n]},y^{''(1)}_{[n]},z_{[n]}|\Hm_{[n]}) \nonumber \\ &\quad + n o(\log P)+ n o(n)  \label{eq:r1r2-ob-1}  \\
					&= I(W_1;y^{'(1)}_{[n]},y^{''(1)}_{[n]},z_{[n]}|\Hm_{[n]}) \nonumber \\ &\quad +I(W_2;y^{'(1)}_{[n]},y^{''(1)}_{[n]},z_{[n]}|\Hm_{[n]}, W_1) \nonumber \\ &\quad + n o(\log P)+ n o(n),  \label{eq:r1-r2} \end{align}
where the transition to~\eqref{eq:r1r2-ob-1} uses the fact that the high SNR variance of $\bar{z}_{t}$ and $z_{t}$ scales as $P^{0}$ and $P^{\alpha_1}$ respectively, which in turn means that knowledge of $\{ y^{'(1)}_{t}, y^{''(1)}_{t},z_{t},\Hm_{[n]}\}_{t=1}^{n}$,
implies knowledge of $W_1,W_2$ and of
$\{x^{1}_{t},x^{2}_{t}\}_{t=1}^{n}$, up to bounded noise level.

Furthermore
\begin{align}
nR_1  \!\!&=\!\! h(W_1) \nonumber \\
			&=\!\! I(\! W_1;y^{'(1)}_{[n]}\!,y^{''(1)}_{[n]}\!,\! z_{[n]}|\Hm_{[n]}\!)\!\!+\!\!h(\! W_1|y^{'(1)}_{[n]}\!,y^{''(1)}_{[n]}\!,\! z_{[n]},\Hm_{[n]}\! )  \nonumber \\
			&= \!\!I(\! W_1;y^{'(1)}_{[n]}\!,y^{''(1)}_{[n]},z_{[n]}|\Hm_{[n]}\! ) \!+\! n o(\log \! P)\!+\! n o(n) , \label{eq:r1}
\end{align}
since again knowledge of
$\{ y^{'(1)}_{t}, y^{''(1)}_{t},z_{t},\Hm_{[n]}\}_{t=1}^{n}$
 provides for $W_1$ up to bounded noise level.

Now combining \eqref{eq:r1-r2} and \eqref{eq:r1}, gives
\begin{align*}
nR_2 \!&=\! I(W_2;y^{'(1)}_{[n]},y^{''(1)}_{[n]},z_{[n]}|\Hm_{[n]},W_1) \!+\! n o(\log P)\!+\! n o(n)  \\
&=I(W_2;y^{'(1)}_{[n]},y^{''(1)}_{[n]}|\Hm_{[n]},W_1) \\
&\quad \!+\! I(W_2;z_{[n]}|y^{'(1)}_{[n]},y^{''(1)}_{[n]},\Hm_{[n]},W_1) \!+\! n o(\log P)\!+\! n o(n) \nonumber\\
&=\! h(y^{'(1)}_{[n]},y^{''(1)}_{[n]}|\Hm_{[n]},W_1)\!-\!\underbrace{h(y^{'(1)}_{[n]},y^{''(1)}_{[n]}|\Hm_{[n]},W_1,W_2)}_{n o(\log P)} \\
&\quad  -\underbrace{h(z_{[n]}|y^{'(1)}_{[n]},y^{''(1)}_{[n]},\Hm_{[n]},W_1,W_2)}_{n o(\log P)} \\
&\quad  +\underbrace{h(z_{[n]}|y^{'(1)}_{[n]},y^{''(1)}_{[n]},\Hm_{[n]},W_1)}_{\le h(z_{[n]})} + n o(\log P)+ n o(n) \\
&\leq \! h(y^{'(1)}_{[n]},y^{''(1)}_{[n]}|\Hm_{[n]},W_1) \!+\!h(z_{[n]})  \!+\! n o(\log P)\!+\! n o(n) \\
&\leq \! h(y^{'(1)}_{[n]},y^{''(1)}_{[n]}|W_1,\Hm_{[n]}) + n \alpha_1\log P \\ &\quad + n o(\log P)+ n o(n),  \label{eq:R2-bound-final}
\end{align*}
which is combined with \eqref{eq:2R1-bound} to give
\begin{align}
2nR_1+nR_2 &\leq 2n\log P+n \alpha_1\log P+ n o(\log P)+ n o(n),
\end{align}
which in turn proves the outer bound
\begin{align}
2d_1+d_2 &\leq 2+\alpha_1,
\end{align}
as described in \eqref{eq:them-outerb-2d1d2}.  Finally interchanging the roles of the two users and of $\alpha_1,\alpha_2$, gives
\begin{align}
d_1+2d_2 &\leq 2+\alpha_2.
\end{align}
Naturally the single antenna constraint gives that $d_1\leq 1, d_2\leq 1$.
\hfill $\Box$


\end{document}